\documentclass[reprint,amsmath,amssymb,aps,prl]{revtex4-2}

\usepackage{graphicx,adjustbox} %Lets us rotate symbols
\usepackage{dcolumn}% Align table columns on decimal point
\usepackage{bm}% bold math
\usepackage[makeroom]{cancel}
\usepackage{url}

\usepackage{breakurl}
\usepackage{graphicx}
\usepackage{xcolor} % add hypertext capabilities
\usepackage{esint} %for slashed integral symbol \fint
\usepackage{xr-hyper}
\usepackage{subfiles}

\usepackage[breaklinks]{hyperref}
\hypersetup{
	colorlinks=true,
	linktoc=page,
	citecolor=americanrose,
	linkcolor=cadmiumgreen,
	urlcolor=blue} 
\def\appendixautorefname~#1\null{Appendix\,#1\null}
\def\sectionautorefname~#1\null{Section\,#1\null}
\def\subsectionautorefname~#1\null{Section\,#1\null}
\def\figureautorefname~#1\null{Figure\,#1\null}
\def\equationautorefname~#1\null{Equation\,(#1)\null}

\definecolor{americanrose}{rgb}{1.0, 0.01, 0.24}
\definecolor{cadmiumgreen}{rgb}{0.0, 0.42, 0.24}

\usepackage[colorinlistoftodos]{todonotes}
\setuptodonotes{color=cyan}
\presetkeys{todonotes}{inline}{}
\reversemarginpar %notes of the left
\makeatletter\g@addto@macro\bfseries{\boldmath}\makeatother

\usepackage[scr=dutchcal,frak=mma]{mathalfa}

\def\braa#1{% \vspace*{4mm} 
        \left\langle\smash{#1}{\vphantom1}\right|}

\def\keta#1{%
        \left|\smash{#1}{\vphantom1}\right\rangle}

\begin{document}

\title{The first law of binary black hole scattering}

\author{Riccardo Gonzo}\email{rgonzo@ed.ac.uk}
\affiliation{Higgs Centre for Theoretical Physics, School of Physics and Astronomy, 
University of Edinburgh, EH9 3FD, UK}%
\author{Jack Lewis}\email{J.E.Lewis@soton.ac.uk} 
\affiliation{School of Mathematical Sciences and STAG Research Centre, University of Southampton, Southampton SO17 1BJ, United Kingdom}%
\author{Adam Pound}\email{A.Pound@soton.ac.uk} 
\affiliation{School of Mathematical Sciences and STAG Research Centre, University of Southampton, Southampton SO17 1BJ, United Kingdom}%

\date{\today}

\begin{abstract}
In the last decade, the first law of binary black hole mechanics played an important unifying role in the gravitational two-body problem. More recently, binary black hole scattering and the application of high-energy physics methods have provided a new avenue into this classical problem. In this Letter, we connect these two themes by extending the first law to the case of scattering orbits. We present derivations based on classical S-matrix, Hamiltonian, and pseudo-Hamiltonian methods, the last of which allows us to include dissipative effects for the first time.  Finally, a ``boundary to bound'' map links this first law to the traditional bound-orbit version. Through this map a little-known observable for scatter orbits, the elapsed proper time, is mapped to the Detweiler redshift for bound orbits, which is an invariant building block in gravitational waveform models.
\end{abstract}

\maketitle

\emph{Introduction.}---The discovery of gravitational waves (GWs) from compact binary systems opened a new chapter in astronomy. Given the enhanced sensitivity and expanded frequency range of future GW detectors, we expect a dramatic increase in the number and variety of detectable compact binary sources~\cite{TianQin:2015yph,Ruan:2018tsw,Reitze:2019iox,Maggiore:2019uih,LISA:2022yao}. Increasingly accurate waveform models will be needed to detect and analyse these sources~\cite{Purrer:2019jcp,Dhani:2024jja}, calling for the development of new tools to study the classical two-body problem. 

Motivated by GW modeling, a host of techniques have been developed to solve the two-body problem in general relativity, including numerical relativity, which numerically solves the fully nonlinear Einstein equations~\cite{Baumgarte-Shapiro}; gravitational self-force (GSF) theory, a perturbative method that applies when one body is much smaller than the other~\cite{Barack:2018yvs}; and Post-Newtonian (PN) and Post-Minkowskian (PM) theory, weak-field expansions that apply when the two bodies are widely separated~\cite{Blanchet:2013haa,Damour:2017zjx}. Historically, focus has been on the bound, inspiralling systems that are the dominant sources for GW detectors. However, the case of hyperbolic, scattering encounters is now of great interest: it is now known that data for scattering orbits can inform bound-orbit models using the effective one-body framework \cite{Buonanno:1998gg,Buonanno:2000ef,Damour:2016gwp,Damour:2017zjx,Khalil:2022ylj,Buonanno:2024vkx,Buonanno:2024byg} or through an analytic continuation from scattering to bound observables~\cite{Kalin:2019rwq,Kalin:2019inp,Saketh:2021sri,Cho:2021arx,Gonzo:2023goe,Adamo:2024oxy}, spurring the development of new particle physics tools~\cite{Bern:2019crd,Bjerrum-Bohr:2022blt,Buonanno:2022pgc,Kosower:2022yvp,DiVecchia:2023frv} that have enabled analytical computations of the two-body scattering Hamiltonian and related observables at high PM order~\cite{Neill:2013wsa,Cheung:2018wkq,Bern:2019nnu,Bern:2021dqo,Bern:2021yeh,DiVecchia:2021bdo,DiVecchia:2021ndb,Bjerrum-Bohr:2021din,Damgaard:2021ipf,Damgaard:2023ttc,Kalin:2020fhe,Dlapa:2021vgp,FebresCordero:2022jts,Dlapa:2022lmu,Dlapa:2024cje,Jakobsen:2022zsx,Jakobsen:2023ndj,Jakobsen:2023hig,Driesse:2024xad,Brandhuber:2021eyq,Cho:2023kux,Akpinar:2024meg}. 

In the bound case, synergies between different methods have consistently helped drive progress~\cite{LeTiec:2014oez,Bini:2019nra}. An important tool in those synergies has been the first law of binary black hole (BH) mechanics~\cite{LeTiec:2011ab,Gralla:2012dm, Blanchet:2012at, LeTiec:2013iux, LeTiec:2015kgg, Fujita:2016igj, Blanchet:2017rcn, Ramond:2022ctc}, which describes how a binary system responds to variations of its parameters (see also \cite{Friedman:2001pf}).  This law has played an important role in the most accurate GSF waveform model~\cite{Wardell:2021fyy,Pound:2019lzj} and in utilizing GSF results within PN, effective one body, and numerical relativity calculations~\cite{Damour:2012ky,Bini:2013zaa}; see~\cite{LISAConsortiumWaveformWorkingGroup:2023arg} for a review. For spinless particles, the binary's response to variations is determined by a basis of observables $\mathcal{B}^{<}$ consisting of the periastron advance $\Delta \Phi$, the radial frequency $\Omega_r$, and the averaged Detweiler redshift $\langle z \rangle$~\cite{LeTiec:2011ab,LeTiec:2015kgg}. 

To date, a first law for scattering scenarios has not been derived. In this Letter, we establish such a law and find the corresponding basis of scattering observables $\mathcal{B}^{>}$. Two of these observables are well studied: the deflection angle $\chi$ and the time delay. %$\Delta t$. 
We complete the basis with a third observable: the elapsed proper time $\Delta \tau$. Our approach is based on a pseudo-Hamiltonian formulation of GSF theory~\cite{Fujita:2016igj,Isoyama:2018sib,Blanco:2022mgd,1SFHamiltonian}. This allows us to include dissipative contributions in the first law, unlike all previous formulations for bound orbits. By comparing to bound-orbit formulations, we also establish a novel analytic continuation between the elements of the scattering $\mathcal{B}^{>}$ and bound $\mathcal{B}^{<}$ bases of observables.

Finally, we link our calculations to high-energy physics methods, proving that the exponential representation of the classical S-matrix \cite{Damgaard:2021ipf,Damgaard:2023ttc} provides a generating functional for the basis of scattering observables and deriving a first law from a PM Hamiltonian.

\paragraph{Conventions} We use geometric units with $G=c=1$ and the $(-+++)$ metric signature.

\emph{First law in the probe limit.}---In the GSF approach, the smaller body (of mass $m_1$) is treated as a point-particle perturbing the spacetime of the large body, which we take to be a Schwarzschild BH of mass $m_2$. We first consider the probe limit (0SF order), in which the particle moves on a geodesic of the Schwarzschild metric $g_{\alpha \beta}^{\text{Schw}}$. The particle's motion is governed by the geodesic Hamiltonian $H_0 = (1/2) g^{\mu \nu}_{\text{Schw}} p_{\mu} p_{\nu}$, where $p_{\alpha} = m_1 g_{\alpha \beta}^{\text{Schw}}\mathrm{d} x^{\beta}/\mathrm{d} \tau$ is the particle's 4-momentum and $\tau$ is its proper time. Assuming, without loss of generality, that the motion lies on the equatorial plane $\theta = \pi/2$, we label the position of the particle with $x^\alpha\left(\tau\right)=\left(t(\tau), r(\tau), \pi/2, \varphi(\tau)\right)$. Because of Schwarzschild's Killing symmetries, the particle's energy and angular momentum $E = -p_{t,0}$ and  $L= p_{\phi,0}$ are conserved (here and below, a subscript 0 indicates the on-shell geodesic value). We now consider unbound geodesic orbits that begin and end at $r=\infty$; such orbits have $E>m_1$ and $L > L_{\text{crit}}(E)$, where $L_{\text{crit}}(E)$ is a critical value of the angular momentum~\cite{Barack:2022pde}.

Following Carter's application of Hamilton-Jacobi theory~\cite{Carter:1968rr}, we use the constants of motion $P_i=(m_1,E,L)$ as canonical momenta and transform to canonical coordinates $(X^i,P_i)$ using the type-2 generating function
\begin{align}
    W(t,r,\varphi;P_i) &= -E t + L\varphi + I_{r,0}(r;P_i)\,, \nonumber \\
    I_{r,0}(r;P_i) &= \int_{r_{\mathrm{m}}}^r \mathrm{d} r \, p_{r,0}(r;P_i)\,,
    \label{eq:canonicalW}
\end{align}
where $r_\mathrm{m}(P_i)$ is the geodesic's minimum radius (i.e., closest approach to the BH). $g^{\mu\nu}_{\rm Schw}p_{\mu,0} p_{\nu,0} = -m_1^2$ implies 
\begin{align}
%    \,, \nonumber \\
    &p_{r,0}(r;P_i) = \frac{\sqrt{E^2 r^4 -r (r-2 m_2) (L^2+m_1^2 r^2)}}{r (r -2 m_2)}\,.
    \label{eq:radial_Schw}
\end{align}
In the coordinates $(X^i,P_i)$, where $X^i=\partial W/\partial P_i$, the Hamiltonian is simply its on-shell value, $H_0=-m_1^2/2$, meaning that Hamilton's equations become~\cite{Schmidt:2002qk}
\begin{align}
    m_1 \frac{\mathrm{d} X^i}{\mathrm{d} \tau}={\frac{\partial H_0}{\partial P_i}} = -m_1\delta^i_1\,.
    \label{eq:gen_coordinates}
\end{align}
Therefore $X^2$ and $X^3$ are constants, while $X^1$ is linear in $\tau$. Since $X^i=\partial W/\partial P_i$, this implies
\begin{align}
 \frac{\partial W}{\partial E} \Bigg|_{\mathrm{out}} &= \frac{\partial W}{\partial E} \Bigg|_{\mathrm{in}}\,,  \qquad \frac{\partial W}{\partial L} \Bigg|_{\mathrm{out}} = \frac{\partial W}{\partial L} \Bigg|_{\mathrm{in}}\,, \nonumber \\
\tau_{\mathrm{out}} - \tau_{\mathrm{in}} &= - \left[\frac{\partial W}{\partial m_1} \Bigg|_{\mathrm{out}} - \frac{\partial W}{\partial m_1} \Bigg|_{\mathrm{in}} \right]\,,
 \label{eq:solution_geo_observables}
\end{align}
where ``in'' and ``out'' denote the initial, incoming state and final, outgoing state. 

Equations~\eqref{eq:canonicalW} and \eqref{eq:solution_geo_observables} imply that the total changes in coordinate time, azimuthal angle, and proper time between initial and final states are
\begin{align}
\label{eq:basis_observables}
  &\hspace{-8pt}\Delta t_0 = t_{\mathrm{out}} - t_{\mathrm{in}} = \frac{\partial}{\partial E} \left[I_{r,0}(r_{\mathrm{out}};P_i) - I_{r,0}(r_{\mathrm{in}};P_i) \right], \\
&\hspace{-8pt}\Delta \varphi_0 = \varphi_{\mathrm{out}} - \varphi_{\mathrm{in}} = -\frac{\partial}{\partial L} \left[I_{r,0}(r_{\mathrm{out}};P_i) - I_{r,0}(r_{\mathrm{in}};P_i) \right], \nonumber \\
 &\hspace{-8pt}\Delta \tau_0 = \tau_{\mathrm{out}} - \tau_{\mathrm{in}} = -\frac{\partial}{\partial m_1} \left[I_{r,0}(r_{\mathrm{out}};P_i) - I_{r,0}(r_{\mathrm{in}};P_i) \right]. \nonumber 
\end{align}

We are interested in the limit where the initial and final states are defined at past and future timelike infinity, with $r_{\rm in}=\infty = r_{\rm out}$, passing through the single radial turning point $r_{\mathrm{m}}$. In this limit, $\Delta\varphi_0$ remains finite, but $I_{r,0}(r_{\rm in/out};P_i)$, $\Delta t_0$, and $\Delta\tau_0$ all diverge. However, we can define regularized versions. Using a convenient dimensionless regulator $\epsilon$~\cite{Gonzo:2023goe}, we first define the scattering radial action 
\begin{align}
    \hspace{-10pt}I^{>,\epsilon}_{r,0}&(P_i) = 2 \int_{r_{\mathrm{m}}}^{+\infty} \mathrm{d} r \, r^{\epsilon} \, p_{r,0}(r;P_i) \,.
    \label{eq:radial_Schw_reg}
\end{align}
Intermediate results depend on the finite value of $\epsilon$, but we obtain $\epsilon$-independent observables in the limit $\epsilon\to0$; when necessary, functions are first defined in regions of the complex-$\epsilon$ plane where integrals converge [e.g., ${\rm Re}(\epsilon)<-1$ in Eq.~\eqref{eq:radial_Schw_reg}] and are then analytically continued to $\epsilon=0$~\footnote{We could equivalently define the radial action as the Hadamard finite part of Eq.~\eqref{eq:radial_Schw_reg}, which is the coefficient of $\epsilon^0$ in the integral's Laurent series around the $\epsilon=0$~\cite{Blanchet:2000nu}. The scattering angle and elapsed times then inherit this Hadamard regularization.}. In terms of $I^{>,\epsilon}_{r,0}$ we can write the regularized $r_{\rm in/out}\to\infty$ limit of Eq.~\eqref{eq:basis_observables} for the full scattering path as
\begin{align}
\Delta\varphi^\epsilon_0 = \pi + \chi^\epsilon_0 &= - \frac{\partial}{\partial L} I^{>,\epsilon}_{r,0}(P_i) \,,
\label{eq:deflection-angle}
\end{align}
where $\chi_0\!=\!\lim_{\epsilon\to0}\chi^\epsilon_0$ is the physical scattering angle, and 
\begin{align}
\Delta t^{\epsilon}_0 = \frac{\partial}{\partial E} I^{>,\epsilon}_{r,0}(P_i)\,, \qquad \Delta \tau^{\epsilon}_0 = - \frac{\partial}{\partial m_1} I^{>,\epsilon}_{r,0}(P_i)\,.
\label{eq:time-def}
\end{align}
Unlike $\chi^\epsilon_0$, the elapsed times $\Delta t^\epsilon_0$ and $\Delta\tau^\epsilon_0$ diverge as $\epsilon\!\to\!0$. The associated physical observables, which are well defined when $\epsilon\!\to\!0$, are \emph{relative} measurements: the difference between $\Delta t_0 (P_i)$ along the geodesic and $\Delta t_0 (P_{i,\text{ref}})$ along some reference orbit; and these relative quantities will take the same values as if we had worked consistently with finite $r_{\rm in/out}$ and only taken the limit $r_{\rm in/out}\to\infty$ at the end of the calculation. In the Supplemental Material~\ref{sec:supp}, we provide exact expressions for the geodesic scattering observables as well as the first few terms in their PM expansions (corresponding to $m_1 m_2 / L\ll 1$). 

Finally, Eqs.~\eqref{eq:deflection-angle} and \eqref{eq:time-def} can be immediately combined into a single equation,
\begin{align}\label{eq:first law 0SF}
   \delta I^{>,\epsilon}_{r,0} = -(\pi + \chi_0) \delta L+ \Delta t^{\epsilon}_0 \delta E - \Delta \tau^{\epsilon}_0\delta m_1\,.
\end{align}
This is our first law for scattering geodesics. Here and below, we discard terms that vanish at $\epsilon=0$, and equalities should be understood in this sense.

\emph{First law at all SF orders.}---Beyond leading order in the mass ratio $m_1/m_2$, the particle generates a metric perturbation $h_{\alpha\beta}$ on the Schwarzschild background. $m_1$ then moves on a geodesic of a certain \emph{effective} metric $\tilde g_{\alpha\beta} = g_{\alpha\beta}^{\text{Schw}} + h^{\rm R}_{\alpha\beta}$~\cite{Poisson:2011nh}, where $h^{\rm R}_{\alpha\beta}$ is a certain regular piece of $h_{\alpha\beta}$. At 1SF order, we can write $h^{\rm R}_{\alpha\beta}$ in terms of the Detweiler-Whiting Green's function $G_{\mathrm{R}}^{\alpha \beta \alpha^{\prime} \beta^{\prime}}$~\cite{Detweiler:2002mi}: %obtained by solving Einstein's equation
\begin{align}
 \hspace{-6pt}   h_{\mathrm{R}}^{\alpha \beta}\!\left(x^\mu;\Gamma\right)= \frac{1}{m_1} \int_\Gamma G_{\mathrm{R}}^{\alpha \beta \alpha^{\prime} \beta^{\prime}}\!\!\left(x^\mu, x'^\mu(\tilde\tau')\right) \tilde{p}_{\alpha^{\prime}}\tilde{p}_{\beta^{\prime}}{\rm d}\tilde\tau'\,.
    \label{eq:hmunu_R}
\end{align}
Here $\tilde{p}_\alpha:=\tilde g_{\alpha \beta} {\rm d}x^\beta/{\rm d}\tilde\tau$, $\tilde\tau$ is proper time in $\tilde g_{\alpha \beta}$, and $\Gamma$ denotes the particle's phase-space trajectory. Due to curvature-induced tail effects, $G_{\mathrm{R}}^{\alpha \beta \alpha^{\prime} \beta^{\prime}}$ is nonzero for all points $x'^\mu$ in the past of $x^\mu$, implying $h_{\mathrm{R}}^{\alpha \beta}$ at a point on $\Gamma$ depends on the entire prior history of $\Gamma$.  At higher SF orders, there is no known Green's-function form analogous to~\eqref{eq:hmunu_R}, but an appropriate $h^{\rm R}_{\alpha\beta}(x^\mu;\Gamma)$ exists at all SF orders~\cite{Pound:2015tma}.   

In this setting, we again consider scattering orbits with initial parameters $P_i=(m_1,E,L)$. The particle's energy and angular momentum evolve due to dissipation, but the orbit remains planar ($\theta=\pi/2$, $\tilde p_\theta=0$). For $L$ above a critical threshold, the orbit remains close to a Schwarzschild geodesic with the same initial $P_i$~\cite{Barack:2022pde,Long:2024ltn}.

Since the motion is geodesic in $\tilde g_{\mu\nu}$, it obeys Hamilton's equations with the test-mass pseudo-Hamiltonian $H = (1/2) \tilde g^{\mu \nu} \tilde{p}_{\mu} \tilde{p}_{\nu}$~\cite{Fujita:2016igj,Isoyama:2018sib,Blanco:2022mgd}. However, we deviate from~\cite{Fujita:2016igj,Isoyama:2018sib} by restricting  to the 4D phase space $(x^A, \tilde p_A)$ satisfying the on-shell condition $H = -m_1^2/2$, with $x^A=(r,\varphi)$. Solving the on-shell condition for $\tilde{p}_t=-\mathcal{H}(t, x^A, \tilde{p}_A;\Gamma)$ gives the new pseudo-Hamiltonian
\begin{align}
\hspace{-6pt}  \mathcal{H}%(t, x^A, \tilde{p}_A;\Gamma)%&=\mathcal{H}_0\left(x_p^i, \tilde{p}_i\right)+ \mathcal{H}_1\left(t, x_p^i, \tilde{p}_i\right) \nonumber \\
  &= \frac{1}{\tilde g^{tt}}\Bigl[\tilde g^{tA}\tilde p_A-\sqrt{(\tilde g^{tA}\tilde p_A)^2-\tilde g^{tt}\bigl(\tilde g^{AB}\tilde p_A \tilde p_B+m_1^2\bigr)}\Bigr]\nonumber\\
  &= -p_{t}(r,\tilde{p}_A) - \frac{h^{\mu \nu}_{\rm R}(t,x^A;\Gamma) \tilde{p}_{\mu} \tilde{p}_{\nu}}{2g^{tt}_{\rm Schw}\,p_{t}}+{\cal O}\!\left(\frac{m_1^2}{m_2^2}\right),\!\!
  \label{eq:Hcal_def}
\end{align}
with $t$ now the parameter along the trajectory. $\mathcal{H}$ is referred to as a pseudo-Hamiltonian because it depends on the trajectory $\Gamma=\left\{\left(x^A(t), \tilde{p}_A(t)\right) \mid t \in \mathbb{R}\right\}$. Hamilton's equations in this context read
\begin{align}
  \frac{{\rm d} x^A}{{\rm d} t}= \left[\frac{\partial \mathcal{H}}{\partial \tilde{p}_A} \right] \quad \text { and } \quad \frac{{\rm d} \tilde{p}_A}{{\rm d} t}=-\left[\frac{\partial \mathcal{H}}{\partial x^A} \right]\,,
  \label{eq:HJ_6D}
\end{align}
where $[\cdot]$ indicates specification of $\Gamma$ as the self-consistent trajectory~\cite{Pound:2009sm} passing through $(x^A,\tilde p_A)$; prior to that specification, $\Gamma$ is treated as independent, and derivatives do not act on it. We go further by replacing $m_1$ by $m_1'$ in Eq.~\eqref{eq:hmunu_R}, setting $m'_1=m_1$ only when $[\cdot]$ is applied. Importantly, Eq.~\eqref{eq:HJ_6D} captures the full dynamics, including dissipation, unlike an ordinary Hamiltonian description.

We derive the first law from ${\cal H}$. Doing so will require its relationship to the redshift $z$: 
\begin{align}\label{eq:z}
     z := \frac{\mathrm{d} \tilde\tau}{\mathrm{d} t} = \left[\frac{\partial \mathcal{H}}{\partial m_1}\right] \,.
\end{align}
To establish this relationship, we consider the normalization condition
\begin{align}
  -m_1=\tilde{p}_\mu \frac{\mathrm{d} x^\mu}{\mathrm{d} \tilde{\tau}}%&=\left(-\mathcal{H}+\tilde{p}_i \frac{d x_p^i}{d t}\right) z^{-1} \nonumber \\
  &=\left(-\mathcal{H}+\tilde{p}_A \left[ \frac{\partial \mathcal{H}}{\partial \tilde{p}_A}\right]  \right) z^{-1} \,,
  \label{eq:norm_redshift}
\end{align}
where we used $\tilde{p}_t = -\mathcal{H}$ together with \eqref{eq:HJ_6D}. Next we note the first line of~\eqref{eq:Hcal_def} shows that, at fixed $m'_1$, $\mathcal{H}$ is a homogeneous function of $\left(m_1,\tilde{p}_A\right)$ of order 1. Euler's homogeneous function theorem hence implies 
$\mathcal{H}= m_1 \frac{\partial \mathcal{H}}{\partial m_1}+\tilde{p}_A \frac{\partial \mathcal{H}}{\partial \tilde{p}_A}$.  Comparing this with~\eqref{eq:norm_redshift}, we obtain~\eqref{eq:z}.

Now, to derive the first law, we loosely follow~\cite{LeTiec:2015kgg} by considering how ${\cal H}$ changes under variations $\delta P_i$ of the initial data, with $\delta x^A_{\rm in}=0$. Given~\eqref{eq:HJ_6D} and \eqref{eq:z}, we find
\begin{align}
      [\delta \mathcal{H}]  &= \left[ \frac{\partial \mathcal{H}}{\partial x^A} \delta x^A \right] + \left[ \frac{\partial \mathcal{H}}{\partial \tilde{p}_A} \delta \tilde{p}_A \right] + \left[ \frac{\partial \mathcal{H}}{\partial m_1} \delta m_1 \right] \nonumber \\
    &= - \frac{\mathrm{d} \tilde{p}_A}{\mathrm{d} t} \delta x^A + \frac{\mathrm{d} x^A}{\mathrm{d} t} \delta \tilde{p}_A + \frac{\mathrm{d} \tilde\tau}{\mathrm{d} t} \, \delta m_1 \,.
    \label{eq:variation_H}
\end{align}
Since $h^{\rm R}_{\mu\nu}$ vanishes for an inertial particle in Minkowski~\cite{Poisson:2011nh}, its contribution to $\tilde p_A$ and ${\cal H}$ vanishes in the initial state, such that $\tilde p^{\rm in}_\varphi=p^{\rm in}_{\varphi,0}=L$ and ${\cal H}_{\rm in}=-p^{\rm in}_{t,0}=E$.  We isolate $\delta L$ and $\delta E$ in~\eqref{eq:variation_H} by defining `interaction' quantities $\tilde {\mathscr{p}}_\varphi:=\tilde p_\varphi-L$, $\tilde{\mathscr{p}}_r=\tilde p_r$, and $\mathscr{H}:={\cal H}-E$, such that
\begin{equation}
  \delta E + [\delta \mathscr{H}] = - \frac{\mathrm{d} \tilde{\mathscr{p}}_A}{\mathrm{d} t} \delta x^A + \frac{\mathrm{d} x^A}{\mathrm{d} t} \delta \tilde{\mathscr{p}}_A + \frac{\mathrm{d} \varphi}{\mathrm{d} t} \delta L + \frac{\mathrm{d} \tilde\tau}{\mathrm{d} t} \delta m_1 \,.
    \label{eq:local_firstlaw}
\end{equation}
Next, we integrate~\eqref{eq:local_firstlaw} along the physical scattering trajectory from $t=-\infty$ to $t=+\infty$, introducing the regularized integral $\langle f\rangle_\Gamma :=\int_\Gamma \mathrm{d}t\, r^\epsilon f$ as in the 0SF case. Integrating the term~$ \frac{\mathrm{d} \tilde{\mathscr{p}}_A}{\mathrm{d} t} \delta x^A$ by parts, we obtain
\begin{multline}
    \label{eq:firstlaw_integrated}
   \Delta t^\epsilon\delta E+\langle [ \delta \mathscr{H} ]  \rangle_\Gamma %= \left\langle - \frac{\mathrm{d} \tilde{p}^{\rm int}_A}{\mathrm{d} t} \delta x^A + \frac{\mathrm{d} x^A}{\mathrm{d} t} \delta \tilde{p}^{\rm int}_A \right\rangle_\Gamma  \\
%   &\hspace{-10pt} \qquad \qquad \qquad + \Delta\varphi\delta L  + \Delta\tilde\tau \delta m_1  \nonumber \\
     \\= \delta  \left\langle \tilde{\mathscr{p}}_A \frac{ \mathrm{d} x^A}{\mathrm{d} t} \right\rangle_{\!\Gamma} + \Delta\varphi^\epsilon \delta L  + \Delta\tilde\tau^\epsilon\delta m_1  \,,\!\!
\end{multline}
where  $\Delta t^\epsilon:=\left\langle 1 \right\rangle_\Gamma$, $\Delta \varphi^\epsilon := \bigl\langle \frac{\mathrm{d} \varphi}{\mathrm{d} t} \bigr\rangle_\Gamma$, and $\Delta \tilde\tau^{\epsilon} := \left\langle \frac{\mathrm{d} \tilde\tau}{\mathrm{d} t} \right \rangle_\Gamma$.
We have discarded boundary terms by choosing ${\rm Re}(\epsilon)$ sufficiently negative and discarded terms that arise from derivatives acting on $r^\epsilon$ because they vanish when analytically continued to $\epsilon=0$. Defining also the regularized `interaction' action
\begin{align}
\hspace{-8pt}   I^{>,\epsilon} := \int_\Gamma {\rm d}t\, r^\epsilon\tilde{\mathscr{p}}_A \frac{ \mathrm{d} x^A}{\mathrm{d} t}=I^{>,\epsilon}_r+\int_\Gamma  \mathrm{d} \varphi\, \tilde{\mathscr{p}}_\varphi \,,
\end{align}
with $I^{>,\epsilon}_r:=\int_\Gamma dr\, r^\epsilon \tilde p_r$, we rewrite~\eqref{eq:firstlaw_integrated} as
\begin{align}
   \Delta t^\epsilon\delta E+\langle [ \delta \mathscr{H}]  \rangle_\Gamma &= \delta I^{>,\epsilon} +  \Delta\varphi^\epsilon \delta L + \Delta \tilde\tau^\epsilon \delta m_1  \,.
    \label{eq:firstlaw_integrated2}
\end{align}

We can rewrite Eq.~\eqref{eq:firstlaw_integrated2} in an alternative form by absorbing $\langle [ \delta \mathscr{H} ]  \rangle_\Gamma $ into `renormalized' variables $\{I^{>,\epsilon}_{\rm ren},E_{\rm ren},L_{\rm ren}\}$. Following~\cite{Fujita:2016igj}'s treatment of the bound case, we define renormalized variables 
\begin{align}
   E_{\rm ren} =  \lambda E\,, \quad L_{\rm ren} =  \lambda L\,,\quad {I}^{>,\epsilon}_{\rm ren} =  \lambda I^{>,\epsilon}\,.   \label{eq:renormalized_actions}
\end{align}
Choosing $\lambda(P_i)$ appropriately to eliminate $\langle [ \delta \mathscr{H} ]  \rangle $ from Eq.~\eqref{eq:firstlaw_integrated2}, we are left with 
\begin{align}
   \delta {I}^{>,\epsilon}_{\rm ren} = -(\pi + \chi^\epsilon) \delta L_{\rm ren} + \Delta t^{\epsilon} \delta E_{\rm ren} - \Delta \tilde\tau^{\epsilon} \delta m_1  \,;
    \label{eq:firstlaw_integrated5}
\end{align}
see the Supplemental Material~\ref{sec:supp} for more details. 

Equation~\eqref{eq:firstlaw_integrated5} is the first law for scattering orbits, valid at all SF orders and including all dissipative effects. To help understand the renormalization of the variables, we observe that the first law defines a sense of conjugacy between variables and observables, just as in the first law of thermodynamics. In that sense, the renormalized variables are the ones conjugate to the physical observables. In Ref.~\cite{Lewis:2025ydo}, we show that in the conservative sector, this sense of conjugacy reduces to the usual sense in Hamiltonian mechanics: the renormalized variables are the true, invariant action variables that are canonically conjugate to the system's action-angles. The need for this renormalization stems from the fact that, as highlighted in Ref.~\cite{Blanco:2022mgd}, if a system can be equivalently described by both a pseudo-Hamiltonian and a Hamiltonian, then variables which are conjugate in one description are not generally conjugate in the other. As a consequence, the momenta $\tilde p_\mu$, from which $E$, $L$, and $I^{>,\epsilon}$ are built, are \emph{not} the canonical momenta in a Hamiltonian description of the conservative sector (i.e., they are not the momenta one would define from a Lagrangian for the conservative sector). We refer to Refs.~\cite{Lewis:2025ydo,1SFHamiltonian} for details.

\emph{From scattering to bound.}---There is a well-known analytic continuation between the deflection angle $\chi$ for unbound orbits and the periastron advance $\Delta\Phi$ for bound orbits, as well as between the scattering and bound radial actions~\cite{Kalin:2019rwq,Kalin:2019inp,Gonzo:2023goe}. Here, using the first laws for unbound and bound motion, we extend these analytic continuations to include all the observables in the scattering and bound bases, $\mathcal{B}^{>} = (\chi,\Delta t^{\epsilon},\Delta \tau^{\epsilon})$ and $\mathcal{B}^{<} = (\Delta \Phi,\Omega_r,\langle z \rangle)$. We limit our analysis to 0SF order, as the analytic continuation for the radial action is not known to be valid beyond 0SF order due to nonlocal-in-time tail effects~\cite{Cho:2021arx,Dlapa:2024cje}.

We write the first law for bound geodesics in terms of the bound radial action,
\begin{align}
    I^{<}_{r,0}&(P_i)  =  2 \int_{r_{-}(P_i)}^{r_{+}(P_i)}\, \mathrm{d} r \, p_{r,0}(r;P_i) \,,
    \label{eq:radial_action_bound}
\end{align}
where $r_\mp$ are the orbit's minimum and maximum radius (i.e., the radii at periapsis and apoapsis).
Following the same arguments as for unbound orbits, one can write the accumulated $\varphi$, $t$, and $\tau$ over a single radial period ($T_{r,0}=2\pi/\Omega_{r,0}$) as derivatives of $I^<_{r,0}$, leading to the first law for bound orbits \cite{LeTiec:2013iux,LeTiec:2015kgg}:
\begin{align}
    \label{eq:first_law_bound}
 \hspace{-8pt}  \delta I^{<}_{r,0} &= -(2\pi + \Delta \Phi_0) \delta L  + \frac{2 \pi}{\Omega_{r,0}} \delta E- \frac{2 \pi}{\Omega_{r,0}}\langle z \rangle_0 \delta m_1\,,
\end{align}
where $\langle z \rangle_0 := (1/ T_{r,0}) \int_0^{T_{r,0}}\mathrm{d}t\,\left(\mathrm{d}\tau /\mathrm{d}t\right)_0$. 

Knowing the scatter-to-bound map for the radial action \cite{Kalin:2019inp,Kalin:2020fhe,Gonzo:2023goe},
\begin{align}
    I^{<}_{r,0}&(P_i)  =  \lim_{\epsilon\to0}[I^{>,\epsilon}_{r,0}(E,L,m_1)-I^{>,\epsilon}_{r,0}(E,-L,m_1)],
\end{align}
and comparing Eq.~\eqref{eq:first_law_bound} to Eq.~\eqref{eq:first law 0SF}, we immediately conclude that there is an analytic continuation between the full set of scattering and bound observables:
\begin{align}
\label{eq:B2B_basis}
   \Delta \Phi_0 &= \chi_0 (E,L,m_1) + \chi_0 (E,-L,m_1)\, , \nonumber\\
   \frac{2 \pi}{\Omega_{r,0}} &= \lim_{\epsilon\to0}[\Delta t^{\epsilon}_0 (E,L,m_1) - \Delta t^{\epsilon}_0 (E,-L,m_1)]\,, \nonumber \\
   \frac{2 \pi \langle z \rangle_0 }{\Omega_{r,0}}  &= \lim_{\epsilon\to0}[\Delta \tau^{\epsilon}_0 (E,L,m_1) - \Delta \tau^{\epsilon}_0 (E,-L,m_1)]\,.
\end{align}
We note that the infrared  divergences in $\Delta t^{\epsilon}_0$ and $\Delta\tau^{\epsilon}_0$ cancel in these expressions because the divergences are independent of $L$; see the Supplemental Material~\ref{sec:supp}.

\emph{First law from the S-matrix.}---In this section, we put our first law in the broader context of the quantum S-matrix description of the classical two-body problem~\cite{Kosower:2018adc}. Given the two-body initial state of well-separated massive point particles $\keta{\Psi_{\text{in}}} = \keta{p_1 p_2}$ of mass $m_1$ and $m_2$, the action of the unitary S-matrix operator,
\begin{equation}
     \hat{S} = \mathcal{T}  \exp\left(-\frac{i}{\hbar} \int \mathrm{d} t \, H_{\text {int}}(t)\right)\,,
     \label{eq:action_Hint}
\end{equation}
describes the time evolution of the state in terms of the interaction Hamiltonian $H_{\text {int}}$, with $\mathcal{T}$ denoting time ordering. The classical two-body scattering dynamics in the asymptotic $\hbar \to 0$ limit is equivalently obtained by evaluating the action, and therefore $H_{\rm int}$, on-shell.

Motivated by \eqref{eq:action_Hint}, we define the exponential representation $\hat{S} = \exp\left(i \hat{N} /\hbar \right)$ \cite{Damgaard:2021ipf,Damgaard:2023ttc}, where $\hat{N}$ is a Hermitian operator. We then study the \emph{real-valued} two-body matrix element 
\begin{align}
    N(\mathbb{E},q,m_1,m_2) &:= \langle p_1' p_2'|\hat{N}|p_1 p_2\rangle \,,
    \label{eq:Noper_matrixelement}
\end{align}
where we defined $\keta{\Psi_{\text{out}}} = \keta{p'_1 p'_2}$, the initial \emph{total} energy $\mathbb{E}$ of both particles, and the exchanged momentum  $q^\mu = p_1'{}^{\mu}-p_1{}^{\mu}=p_2^\mu-p_2'{}^\mu$. To make contact with the incoming total angular momentum $\mathbb{L}= (m_1 m_2 \sqrt{\gamma^2 - 1} b)/\mathbb{E}$, where $\gamma := \frac{p^\mu_1+p_1'^\mu}{2} \frac{p_{2\mu}+p'_{2\mu}}{2}$ and $b$ is the impact parameter in the center of mass (CM) frame,  we perform the Fourier transform
\begin{align}
\label{eq:generating_functional}
    &N^{>,\epsilon}(\mathbb{E},\mathbb{L},\{m_a\}) \\
    & = \frac{1}{4 m_1 m_2 \sqrt{\gamma^2-1}} \int\! \frac{\mathrm{d}^{2+2 \epsilon} q}{(2 \pi)^{2+2 \epsilon}} e^{-i \frac{b(\mathbb{L}) \cdot q}{\hbar}}  N(\mathbb{E},q,\{m_a\}) \,, \nonumber 
\end{align}
using dimensional regularization with $d=4+2 \epsilon$. Infrared, $1/\epsilon$ divergences arise due to the long-range nature of gravity, but their analytic structure is understood ~\cite{Weinberg:1965nx}. 

In complete generality, the expectation value \eqref{eq:Noper_matrixelement} is a function of the kinematic data $(\mathbb{E},\mathbb{L},\{m_a\})$, and its variation in the phase space is 
\begin{align}
\label{eq:first_law_smatrix}
\delta N^{>,\epsilon} = c^\epsilon_{\mathbb{L}}\, \delta \mathbb{L}+ c^{\epsilon}_{\mathbb{E}} \,\delta \mathbb{E} + \sum_{a=1,2} c^{\epsilon}_{m_a} \,\delta m_a \,,
\end{align}
where $c^\epsilon_{\mathbb{L}}$, $c^{\epsilon}_{\mathbb{E}}$, and $c^{\epsilon}_{m_a}$ are gauge-invariant coefficients. 

Using insights from the PM Hamiltonian description~\cite{Cheung:2018wkq} and the relation with the radial action~\cite{Damgaard:2023ttc}, we now identify the coefficients with the observables ${\cal B}^>$. First, by matching the scattering angle $\chi$ in the CM frame, it was shown that the matrix element \eqref{eq:Noper_matrixelement} in the conservative case agrees with the radial action up to a constant proportional to $\mathbb{L}$~\cite{Damgaard:2023ttc}:
\begin{align}
 N^{>,\epsilon} \Big|_{\text{cons}} = \underbrace{\int_{\mathcal{C}_r^{>,\epsilon}} \mathrm{d} r \, \tilde{p}_{r,\text{c.m.}}(r;\mathbb{E},\mathbb{L},\{m_a\})}_{\mathbb{I}^{>,\epsilon}_r} + \pi \mathbb{L}\,,
\end{align}
where $\tilde{p}_{r,\text{c.m.}}$ is the radial relative momentum in the CM frame and $\mathcal{C}_r^{>,\epsilon}$ is the contour of integration for scattering orbits, which implicitly includes a regulator $\epsilon$ inherited from the dimensional regularization. In the Supplemental Material~\ref{sec:supp}, using the PM conservative Hamiltonian and its symmetries in the CM frame \cite{Cheung:2018wkq,Kalin:2019inp}, we then provide a proof of the following conservative first law for the two-body scattering problem,
\begin{align}
\hspace{-6pt}\delta N^{>,\epsilon} =  -\chi \delta \mathbb{L}+ \Delta t^{\epsilon} \delta \mathbb{E} - \sum_{a=1,2} \Delta \tau^{\epsilon}_{a} \delta m_a\,,
\label{eq:first_law_conservative}
\end{align}
where $\Delta \tau^{\epsilon}_{a}$ is the elapsed proper time of particle $a$ and $\Delta t^{\epsilon}$ is the elapsed global time.

If we appeal to $N^{>,\epsilon}=\mathbb{I}^{>,\epsilon}_r +\pi\mathbb{L}$ and restrict to variations with $\delta m_2=0$~\footnote{In the GSF context, we would expect the coefficient of such a variation to be related to the surface gravity of the primary BH~\cite{LeTiec:2011ab,Gralla:2012dm}, while in the PM limit it should asymptote to the elapsed proper time of the BH~\cite{Zimmerman:2016ajr,Albalat:2022lfz}.}, then we see Eq.~\eqref{eq:first_law_conservative} is structurally identical to our previous first law~\eqref{eq:firstlaw_integrated5}. However, the quantities in these laws might differ. Even in the conservative sector, the two-body incoming energy $\mathbb{E}$ and angular momentum $\mathbb{L}$ might not agree with the renormalized one-body, SF counterparts $E_{\rm ren}$ and $L_{\rm ren}$. Moreover, while $N^{>,\epsilon}$ is computed here in the CM frame, GSF calculations might be in the initial rest frame of the heavy BH~\cite{Gralla:2021qaf,Cheung:2023lnj,Cheung:2024jpo} or in any `nearby' frame (including the CM frame); the frame of a GSF calculation is implicitly determined by the choice of gauge for the metric perturbations. However, at 0SF order, we can trivially identify $\mathbb{E}$ with $E + m_2$, $\mathbb{L}$ with $L$, and the CM frame with the heavy BH frame, as the relative radial momentum $\tilde{p}_{r,\text{c.m.}}$ coincides with the geodesic one $p_{r,0}$. Then~\eqref{eq:first_law_conservative} identically matches  \eqref{eq:first law 0SF}, and we have $\chi^\epsilon \to \chi^\epsilon_{0}$, $\Delta t^{\epsilon} \to \Delta t^\epsilon_{0}$, and $\Delta \tau^{\epsilon}_{1} \to \Delta \tau^{\epsilon}_{0}$.  At $n$SF order, the matching with our SF first law is more challenging as the dynamics of the heavy BH (as well as the choice of frame) is encoded in a nontrivial way in the metric perturbations~\cite{Detweiler:2003ci,Bonetto:2021exn}. We leave study of this to future work.

\begin{figure}[t!]
\centering
\includegraphics[width=\columnwidth]{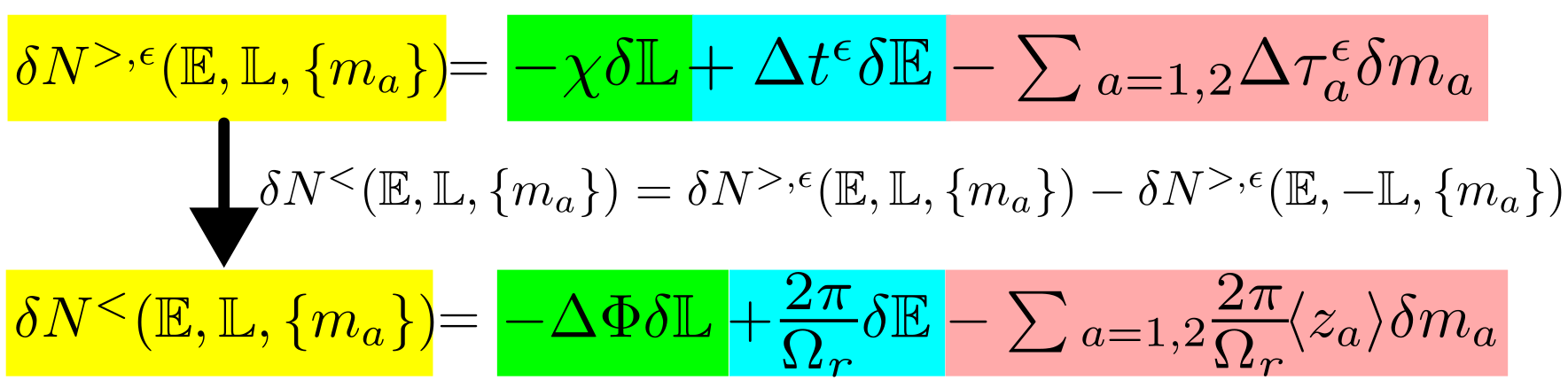}
\caption{Relation between the basis of scattering observables (coordinate time difference $\Delta t^{\epsilon}$, elapsed proper time $\Delta \tau^{\epsilon}$, and deflection angle $\chi$) and the corresponding bound-orbit ones (radial frequency $\Omega_r$, averaged redshift $\langle z \rangle$, and periastron advance $\Delta \Phi$) for the classical gravitational two-body problem.}
\label{fig:firstlaw_B2B}
\end{figure}

Incorporating dissipation in this framework is possible by combining the in-in expectation value \cite{Kosower:2018adc} with the exponential representation of $\hat S$ \cite{Damgaard:2023ttc}: for every observable $\mathcal{O}$~\footnote{Notice that in the limit $\hbar \to 0$, the commutators $ (-i/\hbar)[ \cdot , \cdot ]$ are equivalent to suitable Dirac brackets $\{\cdot,\cdot \}$ which generate the time evolution in the classical phase space~\cite{Gonzo:2024zxo,Kim:2024grz}.},
\begin{align}
 \langle \Delta \mathcal{O} \rangle &= \braa{\Psi_{\text{in}}}  \hat S^{\dagger} \, \mathcal{O} \, \hat S \keta{\Psi_{\text{in}}} - \braa{\Psi_{\text{in}}}  \mathcal{O} \, \keta{\Psi_{\text{in}}} \nonumber \\
 &= \sum_{j=1}^{+\infty} \frac{(-i)^j}{\hbar^j j!}  \underbrace{ [\hat{N},  [\hat{N}, \dots,  [\hat{N}}_{j \text{ times}}, \mathcal{O}] \dots ] ]\,,
\end{align}
where now also the $\hat{N}$ matrix elements with on-shell gravitons are relevant. This suggests a physical principle to connect the coefficients $(c^\epsilon_{\mathbb{L}},c^{\epsilon}_{\mathbb{E}},c^{\epsilon}_{m_a})$ to observables at all orders, with dissipation included; see for example Eq.~(3.48) of~\cite{Damgaard:2023ttc}.

\emph{Conclusion.}---In recent years, the study of unbound orbits through the S-matrix formalism has transformed the gravitational two-body problem. In this Letter, we developed a powerful new tool for such studies: an extension of the first law of binary BH mechanics to the unbound (and dissipative) case. Our derivation at all SF orders utilizes a novel version of the pseudo-Hamiltonian formalism~\cite{Fujita:2016igj,Isoyama:2018sib,1SFHamiltonian}. More generally, we showed how the first law can be derived from the variation of the classical S-matrix in the phase space of the kinematic data $P_i =(\mathcal{E},L,\{m_a\})$; see Fig.~\ref{fig:firstlaw_B2B}. In that context, the S-matrix can be interpreted as a generating functional of classical observables. Among those observables, we have highlighted the elapsed proper time $\Delta \tau^{\epsilon}$ as a new, core element of the (regularized) basis of scattering observables $\mathcal{B}^{>} = (\chi, \Delta t^{\epsilon}, \Delta \tau^{\epsilon})$. 

Using the relation between the scattering and bound radial action, we also established a full correspondence~\eqref{eq:B2B_basis} between the bases of scattering observables $\mathcal{B}^{>}$ and bound observables $\mathcal{B}^{<}$ at 0SF order, as again summarized in Fig.~\ref{fig:firstlaw_B2B}. This extends the well-known map between the deflection angle and periastron advance.

Given the first law's varied applications for bound orbits~\cite{Jaranowski:2012eb,Jaranowski:2013lca,Jaranowski:2015lha,Damour:2014jta,Damour:2016abl,Bernard:2015njp,Bernard:2016wrg,Bernard:2017bvn,LeTiec:2011dp,Akcay:2012ea,Isoyama:2014mja,vandeMeent:2016hel,Barausse:2011dq,Akcay:2012ea,Akcay:2015pjz,Bini:2015bfb,Bini:2016qtx,Bini:2015xua,Colleoni:2015ena,Zimmerman:2016ajr,LeTiec:2017ebm,Barack:2019agd,Pound:2019lzj,Wardell:2021fyy,Albertini:2022rfe}, we expect our work to open many new avenues for scattering calculations. We particularly encourage self-force scattering calculations~\cite{Barack:2022pde,Barack:2023oqp,Whittall:2023xjp,Bini:2024icd,Long:2024ltn} of the observables $\Delta t^\epsilon$ and $\Delta\tilde\tau^\epsilon$. For bound orbits, the 1SF conservative Hamiltonian can be calculated directly from the averaged redshift $\langle z\rangle$, and post-adiabatic waveform models~\cite{Wardell:2021fyy,Pound:2021qin,Mathews:2025nyb} can be written in a gauge-invariant form with $\langle z\rangle$ as an invariant building block~\cite{Lewis:2025ydo,Nasipak:2025tby}. This implies that if the analytic continuation between $\Delta\tilde\tau^\epsilon$ and $\langle z\rangle$ can be extended to 1SF order, then scattering calculations of $\Delta\tilde\tau^\epsilon$ can provide direct inputs to bound-orbit self-force waveform models.

Natural extensions also present themselves. First, one might consider scattering orbits of spinning BHs~\cite{Blanchet:2012at,Ramond:2022ctc,Gonzo:2024zxo}. Second, we considered only two-body matrix elements, but nothing prevents us from studying the variation of matrix elements involving on-shell graviton states, which should be related to the gravitational waveform. Finally, we hope that linking the first laws for scattering and bound orbits beyond 0SF can shed light on the tail effects that have limited the applicability of scatter-to-bound maps~\cite{Cho:2023kux,Dlapa:2024cje}.

\emph{Acknowledgments.}
We thank Alex Le Tiec for comments on a draft of this Letter. RG would like to thank A.~Ilderton, M.~Porrati, K.~Rajeev and C.~Shi for useful discussions. JL and AP gratefully acknowledge helpful conversations with S.~Isoyama and T.~Tanaka about pseudo-Hamiltonian dynamics and with L.~Barack, C.~Kavanagh, R.~Porto, D.~Usseglio, and M.~van de Meent about scatter-to-bound maps. RG is grateful to the Mainz Institute of Theoretical Physics (MITP) of the Cluster of Excellence PRISMA+ (Project ID 390831469), for its hospitality and its partial support during the completion of this work. JL acknowledges the support of an STFC studentship. AP acknowledges the support of a Royal Society University Research Fellowship and a UKRI Frontier Research Grant (as selected by the ERC) under the Horizon Europe Guarantee scheme [grant number EP/Y008251/1].

\bibliographystyle{apsrev4-1_title}
\bibliography{references}% Produces the bibliography via BibTeX.

\clearpage
\appendix 

\renewcommand{\theequation}{S\arabic{equation}}
\renewcommand{\thefigure}{S\arabic{figure}}
\renewcommand{\thesection}{S\arabic{section}}

\onecolumngrid          % often used in APS style
 
\begin{center}
    \Large \textbf{Supplemental Material}
\end{center}
%\vspace{2mm}
\refstepcounter{section}\label{sec:supp}
%\section*{}

In this Supplemental Material, we provide closed-form and PM-expanded formulas for the regularized scattering observables $\mathcal{B}^>=(\chi,\Delta t^\epsilon,\Delta\tau^\epsilon)$ at 0SF order. We then discuss how they relate, through analytic continuation, to the analogous observables for bound geodesics. We also show how the renormalization~\eqref{eq:renormalized_actions} is derived. Finally, we provide a derivation of the conservative first law for scattering orbits from the S-matrix formalism.

\subsection*{Observables at 0SF order}

All of the observables can be obtained from the scattering radial action~\eqref{eq:radial_Schw_reg} via Eqs.~\eqref{eq:deflection-angle} and \eqref{eq:time-def}. Changing variable to $u=1/r$, we write the radial action as
\begin{align}
     I^{>,\epsilon}_{r,0}(P_i) &= 2 \int_{0}^{u_{\rm m}} \frac{\mathrm{d} u}{u^{2+\epsilon}} \,  \frac{\sqrt{B(u)}}{(1 -2 m_2 u)} \,, \nonumber \\
     B(u) &= 2 L^2 m_2 (u-u_{\rm m})(u-u_1)(u-u_2)\,,
     \label{eq:radial_action_u}
\end{align}
where the analytic form of the radial roots $(u_{\rm m},u_1,u_2)$ of $B(u) = 0$ follow from Cartan's formula \cite{Kostic:2012zw}:
\begin{align}
   u_{\rm m} &= \frac{1}{6 m_2} \left(1 +  2 \beta \cos\left(\frac{\zeta-2 \pi}{3}\right)\right)\,,  \quad u_1 = \frac{1}{6 m_2} \left(1 +  2 \beta \cos\left(\frac{\zeta}{3}\right)\right)\,, \nonumber \quad u_2 = \frac{1}{6 m_2} \left(1 +  2 \beta \cos\left(\frac{\zeta+2 \pi}{3}\right)\right)\,,
\end{align}
and we have defined the constants
\begin{align}
\zeta &=\arccos\left(\frac{L^2+\left(36 m_1^2-54 E^2\right) m_2^2}{\beta ^3 L^2}\right)\,, \qquad \beta =\sqrt{1-\frac{12 m_1^2 m_2^2}{L^2}} \,.
\end{align}
 
\emph{Deflection angle.}---The deflection angle $\chi=\Delta\varphi-\pi$ describes the total change of the azimuthal angle in the scattering process, relative to the value ($\pi$) for a straight line in Minkowski spacetime. This quantity does not require any regularization as it is infrared finite. To see how that finiteness follows from the radial action, note that the $1/\epsilon$ divergence of the radial action is \emph{$L$-independent}:
\begin{align}
    I^{>,\epsilon}_{r,0}&(P_i) \Big|_{1/\epsilon} = \frac{2 m_2 \left(2 E^2 -m_1^2\right)}{\sqrt{E^2-m_1^2}} \frac{1}{\epsilon} \,,
    \label{eq:radial_action_div}
\end{align} 
in agreement with amplitude calculations~\cite{Brandhuber:2021eyq}. Therefore, the deflection angle for a scattering geodesic, which is calculated from $\partial I^{>,\epsilon}_{r,0}/\partial L$, is always infrared finite.

Now, to evaluate the deflection angle from \eqref{eq:deflection-angle}, we notice that the action of the $L$-derivative operator on the integration boundary $u_{\rm m}$ vanishes because $B(u_{\rm m}) = 0$, and therefore we only need to act at the integrand level. This allows us to express (see the strategy adopted in~\cite{Gonzo:2023goe}) the geodesic deflection angle as
\begin{align}
\chi_0 = \frac{4 L u_{\rm m}}{\sqrt{E^2-m_1^2}} F_D^{(2)}\left(1,\frac{1}{2},\frac{1}{2},\frac{3}{2}; \frac{u_{\rm m}}{u_1},\frac{u_{\rm m}}{u_2}\right) \,,
\end{align}
where $F_D^{(n)}$ is Lauricella's hypergeometric function
\begin{align}
& F_D^{(n)}\left(\alpha,\left\{\beta_j\right\}_{j=1}^n, \gamma ;\left\{x_j\right\}_{j=1}^n\right)=\frac{\Gamma(\gamma)}{\Gamma(\alpha) \Gamma(\gamma-\alpha)} \int_0^1 \mathrm{~d} t\, t^{\alpha-1}(1-t)^{\gamma-\alpha-1} \prod_{j=1}^n\left(1-x_j t\right)^{-\beta_j} \,.
\label{eq:hypergeometric_FD}
\end{align}

\emph{Coordinate time delay.}---The total change in coordinate time along the geodesic is trivially infinite. Physically it only makes sense to measure the \emph{difference} between this time and the time as measured along a reference path; for an appropriate reference path, that difference is the (Shapiro) time delay~\cite{1964Shapiro}, which represents the amount by which signals are slowed down in the presence of a gravitational source.  Inspired by \cite{Wigner1955,Camanho:2014apa,AccettulliHuber:2020oou}, we take as a reference the clock of an observer moving along a scattering geodesic at very large $L_{\text{ref}} \gg L$ but fixed $E_{\text{ref}} = E$ and define our physical infrared-finite time-delay observable as
\begin{align}
\label{eq:physical_time}
\Delta t^{\text{rel}}_{0,L_{\text{ref}}} = \lim_{\epsilon \to 0} \left[\Delta t^{\epsilon}_0(P_i) - \Delta t^{\epsilon}_0 (P_{i,\text{ref}}) \Big|_{\mathcal{O}\left(\frac{m_1m_2}{L_{\text{ref}}}\right)} \right]\!,
\end{align}
where it is understood that we neglect subleading terms in $m_1 m_2/L_{\rm ref}$. Given \eqref{eq:radial_action_div} and \eqref{eq:time-def}, we notice that this definition is always infrared safe at geodesic order, as expected.  We also note that this definition is completely analogous to $\chi_0$, which we can write as
\begin{align}
&\chi_{0} = \lim_{\epsilon \to 0} \left[\Delta \varphi_0^{\epsilon}(P_i) - \Delta \varphi_0^{\epsilon} (P_{i,\text{ref}}) \right]
\end{align}
because $ \lim_{\epsilon \to 0}\Delta\varphi_0(P_{i,\rm ref})=\pi$.

Using the definition \eqref{eq:time-def} and the representation \eqref{eq:radial_action_u} of the radial action, we can write the regularized elapsed coordinate time as
\begin{align}
&\hspace{-8pt}\Delta t^{\epsilon}_0 = \frac{2 \sqrt{\pi}m_1}{\sqrt{E^2-m_1^2} u_{\rm m}^{1-\epsilon}} \frac{\Gamma(-1+\epsilon)}{\Gamma(-1/2+\epsilon)} F_D^{(3)}\!\left(-1+\epsilon,1,\frac{1}{2},\frac{1}{2},-\frac{1}{2}+\epsilon; 2 m_2 u_{\rm m}, \frac{u_{\rm m}}{u_1},\frac{u_{\rm m}}{u_2}\right),
\end{align}
where we have used the definition \eqref{eq:hypergeometric_FD}. Turning to the physical time delay \eqref{eq:physical_time}, at geodesic order the PM expansion gives the familiar structure~\cite{AccettulliHuber:2020oou,Bautista:2021wfy,Bellazzini:2022wzv,Gonzo:2023goe}:
\begin{align}
\Delta t^{\text{rel}}_{0,L_{\text{ref}}} &= \frac{2 m_2 E \left(3 m_1^2-2 E^2\right)}{\left(E^2-m_1^2\right)^{3/2}} \log\left(\frac{L}{L_{\text{ref}}}\right) + \frac{15 \pi E m_2^2}{2 L} + \mathcal{O}\left(\frac{1}{L^2}\right) \,.
\end{align}

\emph{Relative elapsed proper time.}---The elapsed proper time along scattering orbits has received, to our knowledge, very little attention in the literature (except, possibly, for plunging orbits related to holography applications \cite{Grinberg:2020fdj}). Very recently, ~\cite{Sivaramakrishnan:2024ydy} and \cite{Lee:2024oxo} investigated the proper time from the theoretical and experimental perspective, but without focusing on the two-body problem. In the body of this Letter, we showed that this is a crucial ingredient in our basis of scattering observables. 

Using the same strategy as before, we find that the definition in Eqs.~\eqref{eq:time-def} and the radial action \eqref{eq:radial_action_u} give
\begin{align}
\Delta& \tau^{\epsilon}_0 = \frac{2 m_1 \sqrt{\pi}}{\sqrt{E^2-m_1^2} u_{\rm m}^{1-\epsilon}} \frac{\Gamma(-1+\epsilon)}{\Gamma(-1/2+\epsilon)}  F_D^{(3)}\left(-1+\epsilon,\frac{1}{2},\frac{1}{2},-\frac{1}{2}+\epsilon; \frac{u_{\rm m}}{u_1},\frac{u_{\rm m}}{u_2}\right) \,,
\end{align}
which is the regularized elapsed proper time along scattering orbits. The elapsed proper time relative to the large-$L_{\rm ref}$ reference orbit is, in analogy with Eq.~\eqref{eq:physical_time},
\begin{align}
\hspace{-5pt}\Delta \tau^{\text{rel}}_{0,L_{\text{ref}}} = \lim_{\epsilon \to 0} \left[\Delta \tau^{\epsilon}_0(P_i) - \Delta \tau^{\epsilon}_0 (P_{i,\text{ref}}) \Big|_{\mathcal{O}\left(\frac{m_1m_2}{L_{\text{ref}}}\right)} \right],
\end{align}
which has the PM expansion
\begin{align}
\Delta \tau^{\text{rel}}_{0,L_{\text{ref}}} &= \frac{2 m_1^3 m_2 }{\left(E^2-m_1^2\right)^{3/2}} \log\left(\frac{L}{L_{\text{ref}}}\right) + \frac{3 \pi m_1 m_2^2}{2 L} + \mathcal{O}\left(\frac{1}{L^2}\right) \,.
\end{align}
It is worth noticing that, unlike the time delay  $\Delta t^{\text{rel}}$, the proper time is not defined for massless geodesics and therefore $\Delta \tau^{\text{rel}} \to 0$ as $m_1 \to 0$.

\subsection*{Scatter-to-bound maps at 0SF order}

In the body of the Letter, we derived the scatter-to-bound maps~\eqref{eq:B2B_basis} directly from the first laws for unbound and bound geodesics. Here we observe that these maps also follow from direct inspection of the integral representation of the observables. 

We begin with those representations for unbound geodesics. The observables are simply the (regularized) net change in $\varphi$, $t$, and $\tau$ over the orbit:
\begin{align}
   \pi + \chi_0 = \Delta\varphi_0 &= 2 \int_{r_{\rm m}(P_i)}^{\infty}\, \mathrm{d} r \, \left(\frac{\mathrm{d} \varphi}{\mathrm{d} r}\right)_0\,, \quad \Delta t^\epsilon_0 = 2 \int_{r_{\rm m}(P_i)}^{\infty}\, \mathrm{d} r \, r^\epsilon \left(\frac{\mathrm{d} t}{\mathrm{d} r}\right)_0 \,,  \quad \Delta\tau^\epsilon_0 =  2  \int_{r_{\rm m}(P_i)}^{\infty}\, \mathrm{d} r \, r^\epsilon\left(\frac{\mathrm{d} \tau}{\mathrm{d} r}\right)_0\,.
   \label{eq:unbound_integrals}
\end{align}
We next write the observables for bound orbits in analogous forms. The periastron advance $\Delta \Phi_0$ corresponds to the azimuthal precession over a period of the bound orbit, 
\begin{align}
   2\pi + \Delta \Phi_0 &= 2 \int_{r_{-}(P_i)}^{r_+(P_i)}\, \mathrm{d} r \, \left(\frac{\mathrm{d} \varphi}{\mathrm{d} r}\right)_0\,,
   \label{eq:periastron_advance}
\end{align}
where, recall, $r_{\pm}(P_i)$ are the two radial turning points of the bound geodesic. Similarly, the radial coordinate-time frequency $\Omega_{r,0}$ and the averaged redshift $\langle z \rangle_0$ can be written as~\cite{Damour:1999cr,Barack:2011ed}
\begin{align}
   \frac{2 \pi}{\Omega_{r,0}} &= 2 \int_{r_-(P_i)}^{r_+(P_i)}\, \mathrm{d} r \, \left(\frac{\mathrm{d} t}{\mathrm{d} r}\right)_0 \,,  \quad  \left(\frac{2 \pi}{\Omega_{r,0}}\right) \langle z \rangle_0 =  2  \int_{r_-(P_i)}^{r_+(P_i)}\, \mathrm{d} r \, \left(\frac{\mathrm{d} \tau}{\mathrm{d} r}\right)_0\,.
   \label{eq:redshift_frequency}
\end{align}

Each of the bound-orbit integrals can be written in terms of unbound-orbit integrals by analytically continuing to $E>m_1$ and negative $L$. The key properties in this exercise are
\begin{equation}\label{eq:pr L flip}
p_{r,0}(r;E,L,m_1) = p_{r,0}(r;E,-L,m_1),
\end{equation}
which follows immediately from Eq.~\eqref{eq:radial_Schw}, and the relationship between turning points for bound and unbound  geodesics~\cite{Kalin:2019inp,Kalin:2020fhe,Gonzo:2023goe},
\begin{align}
\label{eq:B2B_roots}
   r_{\pm}(E,L,m_1) = r_{\rm m}(E,\mp L,m_1) \,.
\end{align}
Expressed in terms of $p_{t,0}=-E$, $p_{\varphi,0}=L$, and $p_{r,0}$, the observables read
\begin{align}
   2\pi + \Delta \Phi_0 &= 2 \int_{r_{-}(P_i)}^{r_+(P_i)}\, \mathrm{d} r \, \frac{g^{\varphi\varphi}L}{g^{rr}p_{r,0}(r;P_i)}\,,\nonumber\\
   \frac{2 \pi}{\Omega_{r,0}} &= 2 \int_{r_-(P_i)}^{r_+(P_i)}\, \mathrm{d} r \, \frac{-g^{tt}E}{g^{rr}p_{r,0}(r;P_i)}\,,\nonumber\\
   \left(\frac{2 \pi}{\Omega_{r,0}}\right) \langle z \rangle_0 &= 2  \int_{r_-(P_i)}^{r_+(P_i)}\, \frac{\mathrm{d} r}{g^{rr}p_{r,0}(r;P_i)}\,.\label{eq:bound_integrals}
\end{align} 
In each case, we pull the constant ($E$ or $L$) out of the integral and use Eqs.~\eqref{eq:pr L flip} and \eqref{eq:B2B_roots} to write the integral as
\begin{align}
	 \int_{r_{-}(P_i)}^{r_+(P_i)} \frac{F(r)\mathrm{d} r}{p_{r,0}(r;P_i)}  &= \int_{r_{-}(P_i)}^\infty \frac{F(r)\mathrm{d} r}{p_{r,0}(r;P_i)}   - \int_{r_{+}(P_i)}^\infty \frac{F(r)\mathrm{d} r}{p_{r,0}(r;P_i)} \nonumber\\
	 &= \int_{r_{m}(P_i)}^\infty \frac{F(r)\mathrm{d} r}{p_{r,0}(r;P_i)}  - \int_{r_{\rm m}(E,-L,m_1)}^\infty \frac{F(r)\mathrm{d} r}{p_{r,0}(r;E,-L,m_1)}.  \nonumber 
\end{align}
Here $F(r)$ is one of $g^{\varphi\varphi}/g^{rr}$, $g^{tt}/g^{rr}$, or $1/g^{rr}$, which in each case depends only on $r$. 
Now comparing the observables ~\eqref{eq:bound_integrals} to the observables~\eqref{eq:unbound_integrals} for unbound orbits, and noting Eq.~\eqref{eq:radial_action_div}, we recover Eq.~\eqref{eq:B2B_basis}.

\subsection*{Renormalized action variables in the pseudo-Hamiltonian formulation}

We show here that, choosing an appropriate renormalization $\lambda$ of the action variables, we can write the first law for the pseudo-Hamiltonian \eqref{eq:firstlaw_integrated2} in the convenient form \eqref{eq:firstlaw_integrated5}. We begin with the rescaling defined in \eqref{eq:renormalized_actions},
\begin{align}
   E_{\rm ren} =  \lambda E\,, \quad L_{\rm ren} =  \lambda L\,,\quad {I}^{>,\epsilon}_{\rm ren} =  \lambda I^{>,\epsilon}\,,
\end{align}
which gives the first law in the form
\begin{align}
&\frac{\delta \lambda}{\lambda^2} \left[{I}^{>,\epsilon}_{\rm ren} + (\pi + \chi^\epsilon) L_{\rm ren} - \Delta t^{\epsilon}  E_{\rm ren}\right] + \langle [ \delta \mathscr{H} ]  \rangle_\Gamma = \frac{1}{\lambda} \left[\delta {I}^{>,\epsilon}_{\rm ren} - \Delta t^{\epsilon} \delta E_{\rm ren} + (\pi + \chi^\epsilon) \delta L_{\rm ren} \right] + \Delta \tilde\tau^{\epsilon} \delta m_1\,.
\label{eq:first_law_massaged}
\end{align}
To simplify this equation we appeal to the normalization condition~\eqref{eq:norm_redshift} in the form $m_1 z = -\tilde p_\mu \frac{dx^\mu}{dt}$, which we integrate to obtain 
\begin{equation}
m_1\Delta\tilde\tau^\epsilon = \left\langle[\mathscr{H}]\right\rangle_\Gamma + \Delta t^\epsilon E  -(\pi+\chi^\epsilon) L - I^{>,\epsilon}.
\end{equation}
We then see that Eq.~\eqref{eq:first_law_massaged} reduces to the renormalized first law 
\begin{align}
   \delta {I}^{>,\epsilon}_{\rm ren} = -(\pi + \chi^\epsilon) \delta L_{\rm ren} + \Delta t^{\epsilon} \delta E_{\rm ren} - \Delta \tilde\tau^{\epsilon} \delta m_1  \,
   \label{eq:first_law_ren}
\end{align}
if $\lambda$ satisfies 
\begin{align}
\lambda \langle [ \delta \mathscr{H} ]  \rangle_\Gamma + \langle [ \mathscr{H} ]  \rangle_\Gamma \delta\lambda = \Delta \tilde\tau^{\epsilon} (\lambda - 1) \delta m_1 +  \Delta \tilde\tau^{\epsilon} m_1 \delta \lambda\,.
\label{eq:first_law_massaged2}
\end{align}
If we now expand $\lambda$ as
\begin{align}
\lambda = 1 + \Lambda\,, \qquad  \Lambda = \mathcal{O}(m_1/m_2)\,,
\end{align}
we then obtain from \eqref{eq:first_law_massaged2} that the following condition needs to be satisfied:
\begin{align}
\Delta \tilde\tau^{\epsilon} \delta (m_1 \Lambda) = \langle [ \delta \mathscr{H} ] \rangle_\Gamma + \mathcal{O}(m_1^3/m_2)\,.
\end{align}
We will further investigate this equation, and its solution, in the followup paper~\cite{1SFHamiltonian}.

\subsection*{Conservative first law from the S-matrix formalism}

We prove here the validity of the first law in the conservative case for the S-matrix, combining the relation of the PM Hamiltonian \cite{Neill:2013wsa,Cheung:2018wkq} with amplitudes and with the radial action \cite{Bern:2019crd,Kalin:2019rwq}, which is then linked directly to the exponential representation of the S-matrix \cite{Damgaard:2021ipf,Damgaard:2023ttc}. Albeit the Hamiltonian extracted from the amplitude is gauge-dependent, the following derivation---and therefore our first law---is independent of the choice of gauge.

We first define the center of mass frame for the massive spinless particles 1 and 2 as the one where the incoming states have momenta $\left(E_1, \vec{p}\right)$ and $\left(E_2,-\vec{p}\right)$ while the outgoing states have momenta $\left(E_1, \vec{p}^{\prime}\right)$ and $\left(E_2,-\vec{p}^{\prime}\right)$, with $E_j = \sqrt{|\vec{p}|^2+m_j^2}$. Conservation of energy then implies $|\vec{p}| = |\vec{p}^{\prime}|^2$, and the exchanged momentum in the scattering process becomes $q^{\mu} = (0 ,\vec{q}) = (0 ,\vec{p} - \vec{p}^{\prime})$.

We then introduce polar coordinates in the plane of the motion, $\vec{x} = (r \cos(\phi),r \sin(\phi),0)$, where $r$ is the distance between the two bodies and $\phi$ is the azimuthal phase. We can then define an effective conservative Hamiltonian in the CM frame, 
\begin{align}
	 H^{\text{PM}}_{\text{c.m.}}\left(r, |\vec{p}|\right)=\sum_{a=1,2} \sqrt{|\vec{p}|^2+m_a^2}+V\left(r, |\vec{p}|\right) \,,
\end{align}
where $V\left(r, |\vec{p}|^2\right)$ is the interaction potential for spinless particles, which can be extracted systematically from conservative amplitudes order by order in the PM expansion \cite{Cheung:2018wkq}. Here $x^i$ and $p_i$ are canonically conjugate to each other \cite{Bern:2019crd}, and denoting the norm of the center of mass spatial momentum at infinity $|\vec{p}|(r \to +\infty)$ as 
\begin{align}
	 \hspace{-7pt}p_{\infty} = \frac{1}{4 \mathbb{E}^2} (\mathbb{E}^2-\left(m_1-m_2\right)^2)(\mathbb{E}^2-\left(m_1+m_2\right)^2)\,,
\end{align}
we obtain the conservation laws for the energy and angular momentum, 
\begin{align}
	 \hspace{-8pt}\mathbb{L} = p_{\phi} = b \, p_{\infty} \,, \quad \mathbb{E} = H\!\left(r,|\vec{p}|^2= p_{r,\text{c.m.}}^2+\frac{L^2}{r^2}\right).\!
\end{align}
At this point, we can proceed as in \cite{LeTiec:2015kgg} and consider the variation of our Hamiltonian $H^{\text{PM}}_{\text{c.m.}}(r,p_{r,\text{c.m.}},\phi,\mathbb{L},\{m_a\})$ in the phase space:
\begin{align}
      \delta H^{\text{PM}}_{\text{c.m.}}  &= \frac{\partial H^{\text{PM}}_{\text{c.m.}}}{\partial r} \delta r+\frac{\partial H^{\text{PM}}_{\text{c.m.}}}{\partial p_{r,\text{c.m.}} } \delta p_{r,\text{c.m.}} +\frac{\partial H^{\text{PM}}_{\text{c.m.}} }{\partial \mathbb{L}} \delta \mathbb{L}  + \sum_{a=1,2}  \frac{\partial H^{\text{PM}}_{\text{c.m.}}}{\partial m_a} \delta m_a \nonumber \\
    &= -\frac{\mathrm{d} p_{r,\text{c.m.}} }{\mathrm{d} t} \delta r+\frac{\mathrm{d} r}{\mathrm{d} t} \delta p_{r,\text{c.m.}} +\frac{\mathrm{d} \phi}{\mathrm{d} t} \delta \mathbb{L}  + \sum_{a=1,2}  \frac{\partial H^{\text{PM}}_{\text{c.m.}}}{\partial m_a} \delta m_a\,,
    \label{eq:variation_H_effPM}
\end{align}
where we have used Hamilton's equations in the last line. We can now recognize the redshift variables $z_1$,$z_2$ \cite{Blanchet:2012at},
\begin{align}
    z_{a}  = \frac{\partial H^{\text{PM}}_{\text{c.m.}}}{\partial m_a} = \frac{\mathrm{d} \tau_{a}}{\mathrm{d} t}\,,
\end{align}
where $\tau_a$ is the elapsed proper time of particle $a$ in the CM frame. We can now proceed as in our pseudo-Hamiltonian derivation, using the regularized integration
\begin{align}
  \langle F(\cdot) \rangle  = \int \mathrm{d} t \, r^{\epsilon} F(\cdot) 
\end{align}
on both sides of the equality \eqref{eq:variation_H_effPM}. In this case, though, 
\begin{align}
     \langle H(r,p_r,\phi,\mathbb{L},\{m_a\}) \rangle = \mathbb{E} \, \langle 1\rangle = \mathbb{E} \, \Delta t^{\epsilon}\,,
\end{align}
where $\Delta t^{\epsilon}$ is the elapsed global time. Therefore, following similar steps to the ones in the main text and dropping total boundary terms (which are zero on-shell), we obtain
\begin{align}
\delta \mathbb{I}_r^{>,\epsilon} &=  -(\pi + \chi) \,\delta \mathbb{L}+ \Delta t^{\epsilon} \delta \mathbb{E} - \sum_{a=1,2} \Delta \tau^{\epsilon}_{a} \delta m_a\,,
\label{eq:first_law_conservative_Hamiltonian}
\end{align}
where we have defined
\begin{align}
\mathbb{I}_r^{>,\epsilon}  = \left\langle p_{r,\text{c.m.}} \frac{\mathrm{d} r}{\mathrm{d} t} \right\rangle  \,, \quad \chi = \left\langle \frac{\mathrm{d} \phi}{\mathrm{d} t} \right\rangle \,, \quad \Delta \tau^{\epsilon}_{a} = \left\langle \frac{\mathrm{d} \tau_{a}}{\mathrm{d} t} \right\rangle\,.
\end{align}
Finally, we need to link the radial action defined here with the expectation value of the $\hat{N}$ operator for the two-body conservative case. We now recall the proof in Sec.~3.1.3 of \cite{Damgaard:2023ttc}, where it was shown (non-perturbatively) that  
\begin{align}
\chi  = - \frac{\partial N^{>,\epsilon}(\mathbb{E},\mathbb{L},\{m_a\})}{\partial \mathbb{L} }\,,
\end{align}
where all the other variables are kept fixed; showing therefore that, as expected, 
\begin{align}
 N^{>,\epsilon}  = \mathbb{I}_r^{>,\epsilon} + \pi \mathbb{L} \,.
\end{align}
This concludes the proof of the conservative first law \eqref{eq:first_law_conservative} in the S-matrix formalism.

%\clearpage

%\bibliography{references}

\end{document}